\documentclass{article}
\usepackage{amsmath}
\usepackage{amssymb}

\newtheorem{theorem}{Theorem}
\newtheorem{acknowledgement}[theorem]{Acknowledgement}

\input{tcilatex}
\begin{document}

\title{A Hilbert Space setting for s $\geq $1 interactions which replaces
Gauge Theory}
\author{Bert Schroer \\
present address: CBPF, Rua Dr. Xavier Sigaud 150, \\
22290-180 Rio de Janeiro, Brazil\\
permanent address: Institut f\"{u}r Theoretische Physik\\
FU-Berlin, Arnimallee 14, 14195 Berlin, Germany}
\date{October 2014}
\maketitle

\begin{abstract}
The recently discovered Hilbert space description of renormalizable
interactions of higher spin s$\geq $1 fields requires to replace the
pointlocal s=1 vectorpotentials of indefinite metric (Krein space) BRST
gauge theory by their stringlike counterpart in Hilbert space. It is shown
that the Hilbert space positivity leads to new properties outside the
conceptual range of the gauge theoretic description: topological aspects of
Wilson loops, induced normalization terms, in particular Mexican hat type
potentials for massive vectormesons coupled to Hermitian scalar field and a
possible role of string-localization in confinerment and "darkness".
\end{abstract}

\section{Introduction}

There is no property of quantum theory which is more important than the
positivity coming from its operator formulation in Hilbert space. Born's
probability interpretation of quantum theory (QT) depends on it, and in
quantum field theory (QFT) it is a direct consequence (without invoking Born
definition) of modular localization theory together with the fact that all
physical (i.e. finite energy) states are KMS states after restricting them
to modular localized subalgebras. Hence in QFT there is a direct relation of
quantum causal localization in Hilbert space with statistical ensemble
probability (the ensemble of observables localized in a spactime region)
with which Einstein had no problems \cite{hol} \cite{Sigma}.

Yet gauge theory starts from covariant \textit{pointlike} vectorpotentials
which are incompatible with a Hilbert space and rather act in a suitably
defined indefinite metric Krein space; it is not the linear structure of
Hilbert space but rather the for quantum theory indispensable nonlinear
positivity aspects which is violated in gauge theory and only partially
recovered in terms of gauge invariance. Whereas for free fields it is
trivial to recover the associated pointlike field strengths whose
application to the vacuum state generate a Hilbert space, in the presence of
interactions the indefinite metric becomes deeply enmeshed with the matter
fields (which in zero order are free fields in Hilbert space), so that one
needs a rather elaborate operator gauge formalism which requires an
extension of the Krein space setting by "ghost operators") in order to be
able to extract a subset of local observable fields which act in a Hilbert
space. In this gauge formalism important physical states (e.g. charged
states) remain outside the formalism.

Gauge theory describes the vacuum sector i.e. the Hilbert suspace generated
by the gauge invariant observables acting on the vacuum but the
gauge-variant field (which includes the charge-carrying matter fields of
QED) have no physical meaning. The BRST gauge formalism is a consistent
combinatorial perturbative formalism which is (apart from the mentioned
vacuum setor) outside the functional analytic operator control of quantum
theoy. It is somewhat of a miracle that the extension of the BRST $s$%
-invariance to global objects as scattering process leads to results which
are not only consistent with the formalism but also pass many observational
tests. In the words of Raymond Stora, the main protagonist of BRst gauge
theory "gauge theory is a miracle and one does not understand why, as far as
we know it, the formal rules seem to cover observational results of particle
physics". Most physicists from the older generation (including myself) knew
that quantum gauge theory is a successful "placeholder" for a still unkown
perturbative renormalization theory for $s\geq 1.$ Stanley Mandelstam and
Bryce DeWitt attempted to go beyond gauge theory at a time when important
concepts were still missing.

\textit{The lack of Hilbert space positivity is a gaping wound in the QFT
description of }$s\geq 1~$\textit{interactions and in particular of }$s=1~$%
\textit{gauge theory. It is the principle aim of the present paper to
overcome this limitations and obtain new results and interpretations which
gauge theory misses, as correct description of the charge screening
properties of interacting massive vectormesons in particular in couplings
with Hermitian fields (the Higgs model).}

This is very different from the situation in classical gauge theory where
pointlike vectorpotentials have a well-defined conceptual status as
classical fields and gauge transformations transform classical
vectorpotentials into others in such a way that the observable field
strengths remain invariant. Vectorpotentials are useful classical objects
even though they do not explicitly appear in Maxwell's equations. In
contrast to QFT, Hilbert space positivity and probability have no conceptual
counterpart in classical theory, and hence it is not surprising that the
alleged quantization parallelism between classical and quantum field
theories on which Lagrangian quantization is based fails precisely for $%
s\geq 1~$QFTs as the result of a clash of the pointlike nature of the field
description with the Hilbert space positivity of quantum theory. This
problem is not limited to massless $s\geq 1~$potentials where it is
reflected in the impossibility to obtain pointlike covariant
vectorpotentials in the $s=1$ from the covariantization of Wigner's unitary
positive energy representations of the Poincar\'{e} group.

It \ also shows up in the nonrenormalizabilty of pointlike higher spin
interactions; with other words behind pointlike nonrenormalizability of $%
s\geq 1$ interactions there hides a weakening of localization in such a way
that the interaction becomes renormalizable if one works with stringlocal
fields. The aforementioned gauge formalism is a formal trick to uphold
pointlike localization at the expense of sacrificing Hilbert space
positivity which at the end of the calculations. So quantum gauge theory
comes at a high price in that the gauge invariant part only covers a small
part of the full QFT although, as mentioned before, the applixation of the
gauge formalism outside the vacuum sector leads to unmerited and (quoting
Stora) not understood successes. It turns out that the renormalizable
stringlocal matter fields can be used to define extremely singular
pointlocal fields whose correlation functions are unbounded in momentum
space (infinite $d_{sd}~$in $x$-space); however for massless vectormesons
only the stringlocal matter fields survive; in that case the strings are
rigid \cite{Bu} and different directions are not unitarily equivalent
(Lorentz covariance is spontaneously broken \cite{Froe}) which manifests
itself in terms of the perturbative logarithmic infrared divergencies.

This limitation affects QED, where the absence of a pointlocal physical
electron operator requires to substitute missing spacetime derivations of
collision processes by momentum space prescriptions in terms of
photon-inclusive cross sections for the scattering of charged particles. In
order to improve the conceptual understanding of QFT there are two
alternatives: either construct the charged sectors via representation
theoretical concepts from the vacuum sector, or find a Hilbert space
alternative to the Krein space gauge theory. The first strategy is
successful for QFT with mass gaps\footnote{%
The DHR reconstruction of the superselected sectors \cite{Haag} and their
amalgamation into a field algebra with an inner symmetry \cite{Do-Ro} for
which the fields are not necessarily pointlike but could be semi-infinite
stringlike.}, but presents difficult and largely unsolved problems in case
of the (gauge-invariant) vacuum sector in zero mass gauge theories \cite%
{Bu-Ro}. The second strategy is to realize that the clash between
localization and the Hilbert space structure can also be resolved by ceding
on the side of localization; this causes no conceptual problems since
pointlike localization in a Krein space outside a Hilbert space setting is
anyhow a fake (if used outside the vacuum sector).

It turns out that the conceptual price is rather small compared with what
one may have expected considering the fact that Hilbert space positivity is
a very strong restriction. It simply consists in allowing fields localized
on semi-infinite spacelike strings (in addition to those representing
pointlike observables). However this requires the elaboration of a new
formalism of renormalized perturbation theory which in case of $s=1$
replaces the gauge setting. The basic ideas and their application to second
order perturbation theory, as well as their relevance for the future
development of the Standard Model, is the main theme of the present note
(see also \cite{stringlocal}, \cite{Sigma}). A systematic study of the
problems encountered in the generalization of the Epstein-Glaser causal
approach is in preparation \cite{Jens1} \cite{Jens2} whereas some
preparatory remarks about the new setting can already be found in \cite{MSY} 
\cite{Rio} \cite{Bros} \cite{stringlocal}.

The recognition of the conceptual origin of the problem is not new. Already
Wigner in his 1939 representation theory \cite{Wig} of $m=0$ finite helicity
representations of the Poincar\'{e} group knew that covariant
vectorpotentials which are associated to $(1/2,1/2)$ representations of the
Lorentz-group do not occur in the list of possibilities which arises from
the covariantization of the unique ($m=0,\left\vert h\right\vert =1$) Wigner
representation. This permits an immediate generalization to $s>1$
tensorpotentials; there is no such problem with interaction-free pointlike
massive potentials with short distance dimension $d=s+1;$ they do not allow
to take massless pointlike limits (the Proca potential,..) although this
problem disappears by passing from potentials to field strengths. Zero mass
limits however do exist after converting pointlike potentials into their
covariant stringlike siblings which are associated with the same Wigner
representation as the field strengths; there is no other way for maintaing
the Hilbert space positivity and the standard relation with pointlike field
strengths. All these observation on interaction-free potentials are
well-known.

In the case of couplings involving \textit{massive} $s\geq 1$ fields, the
problem between pointlike fields and the Hilbert space positivity is more
subtle; in this case \textit{the Hilbert space structure clashes with
renormalizability}. The remedy is to convert the pointlike interaction into
its stringlike analog and show that the renormalized perturbation theory
involving covariant fields localized on semi-infinite spacelike strings is
well-defined. The reason why this could work is that stringlocal potentials
have $d=1$ independent of spin, which permits to construct renormalizable
couplings in the sense of power counting ($d_{int}\leq 4$)~for any spin%
\footnote{%
In the present note the spin is assumed to be integer so the the potentials
are bosonic.} in such a way that the local observables will remain pointlike
(possibly pointlike composites of stringlocal fields). The constructions in
this note will be limited to $s=1;~$this is not only because in that case
the new formalism has its simplest realization, but also since it permits
confrontations of new results with observational physics; indeed it is the
first contact of ideas coming from local quantum physics (LQP is the
algebraic approach to QFT \cite{Haag}) with observational properties of the
Standard Model with many new theoretical aspects and new ways of explaining
experimental observations.

Besides providing a basis for the extension of proofs of structural theorems
(spin\&statistics, TCP, the LSZ scattering theory from mass gaps,...) for
which the Hilbert space positivity of quantum theory is essential and which
therefore is not possible in a gauge setting, the existence of physical
matter fields and massive Y-M gluons in the new formulation provides for the
first time a QFT setting for \textit{how confinement can be related to the
infrared divergences of physical fields} in the massless limit. In all those
structural properties, the Hilbert space positivity of the
operator-algebraic setting (not taken care of in functional settings) plays
an essential role, which explains why, apart from the vacuum sector, they
are outside the range of gauge theory. Gauge theory is basically a
perturbative combinatorial structure to which operator methods which rely on
positivity (spectral representations, functional analysis, operator
algebraic methods) cannot be applied.

In the existing literature one finds the observation that the so-called
axial gauge is consistent with Hilbert space positivity. But what prevented
to use this observation as a start of a Hilbert space formulation of $s=1\ $%
interactions is the fact that only the interpretation of this unit vector $%
e\ $as a fluctuating spacelike direction of a field $A_{\mu }(x,e)$
localized on the spacelike line $x+\mathbb{R}_{+}e,$ $e^{2}\equiv e^{\mu
}e_{\mu }=-1~$leads to a consistent formulation. In this new setting every
field has its independent fluctuating $e$-variable, just as the fluctuating $%
x~$which marks the start of a spacelike half-line. This requires in
particular that a Lorentz transformation covariantly changes $e$ (which
contradicts the gauge-parameter interpretation). The non-covariant axial
gauge setting finally fell into disgrace since it leads to confusing
entangled ultraviolet-infrared divergence problems for which no solution was
found. It turns out that it is precisely the directional fluctuation
property in $e$ which reduces the short distance dimension $d=2$ of the
pointlike Proca field $A_{\mu }^{P}$ to $d=1$ of the stringlocal $A_{\mu
}(x,e)$ and thus render the interaction fit for renormalization since it
lowers the short distance dimension of e.g. massive pointlike QED from $%
d_{int}^{P}=5$ to the power-counting compatible value $d_{int}^{S}=4~$for
the stringlocal interaction. \ 

In fact the stringlocal construction of the S-matrix suggests a round-about
definition of higher order pointlike interaction densities whose short
distance behavior is precisely that expected from the unlimited increase of
the momentum space polynomial degree; but in contrast to the direct
pointlike setting, which leads to an ever increasing (with perturbative
order) number of coupling parameters, the coupling strengths are those of
the interaction-defining first order. Although the off-shell correlations of
pointlike fields are expected to have short distance dimensions which
increase with the perturbative order, this is not the case for the on-shell
scattering amplitudes, as will be explained in these notes. In order to
avoid conceptual confusions it is important to point out that the
string-localization refers to fields and not to particles. Particles remain
Wigner particles, and with the exception of the noncompact localized
continuous spin Wigner representation spaces all particle spaces remain
compact localizable.

The new Hilbert space setting has interesting consequences for the Standard
Model. On the one hand, as already mentioned before, the Hilbert space
nature of stringlocal massive Y-M fields (massive gluons) instead of
pointlike unphysical Y-M fields in Krein space opens the possibility of an
understanding of confinement in the limit of vanishing vectormeson mass. By
applying resummation techniques to leading infrared logs\footnote{%
Similar to the proof \cite{YFS} of the vanishing of scattering amplitudes of
charged particles in the presence of only a finite number of photons (in
terms of perturbative resummation techniques).} there are good reasons to
expect, that by using the vectormeson mass as a natural covariant infrared
regulator parameter, the $m\rightarrow 0$ limit vanishes for all
correlations which contain besides an arbitrary number of pointlike
composites also a stringlike gluon (or "$e$-unbridged" $q-\bar{q}$ pairs);
this is the only known interpretation of the meaning of confinement which
explains the non-observability of gluons/quarks and at the same time is
consistent with the foundational causal localization principle (in a Hilbert
space setting!) of QFT. On the other hand it also turns the "Higgs
mechanism" from its head to its feet by realizing that the physical content
of the Higgs model is nothing else than the renormalizable interaction of 
\textit{massive vectormesons coupled to Hermitian} (instead of charged) 
\textit{fields} $H$. In the correct description there is no
symmetry-breaking and \textit{the Mexican hat potential is not put in, but
is rather induced} in second order from the renormalizable stringlocal
reformulation of a pointlike $gA^{P}A^{P}H$ interaction; hence it is not
surprising that the numerical coefficients of the induced potential depend
on the ratio of the masses of the two fields. There is no place for
couplings of massive vectormesons for symmetry-breaking (which symmetry?%
\footnote{%
Local gauge symmetry is not a symmetry but a formalism which permits to
extract physics fom a Krein space decription (although it is in itself
unphysical it leads to a physical Hilbert space subalgebra).}) by
nonvanishing one-point functions of gauge-variant fields; the conceptual
difference between an induced Mexican hat potential and one put into the
interaction in order to support the incorrect idea of mass creation through
spontaneous symmetry breaking in the massless two-parametric scalar QED
cannot be bridged by arguments which are consistent with QED and also not
with the principles of QFT where masses of the model-defining elementary
fields belong to the input data and only boundstate masses of particles
interpolated by composite fields are predictions of the model.

The Hermitian model shares with its complex counterpart ("massive QED") the
screening of its Maxwell charge. Since there is no particle/antiparticle
counting charge, the characteristic property of the Hermitian coupling is
the screened charge of the Maxwell-current which is the only conserved
current of the abelian Higgs model. A conserved current of a spontaneous
broken symmetry leads to a divergent charge (this is really the intrinsic
definition of a spontaneously broken symmetry) whose large distance
divergence is caused by the presence of a massless Goldstone boson (this is
the content of Goldstone's theorem).

The stringlocal interaction associated to the pointlike $gA^{P}A^{P}H$
brings a stringlocal selfadjoint scalar $\phi (x,e)$ into the game which
(unlike the Higgs field $H$) shares its degrees of freedom with those of the
massive vectorpotentials i.e. does not result from an additional coupling;
the degrees of freedom are not changed by the presence of the "intrinsic
escort" field $\phi $. The appearance of these stringlocal intrinsic escorts
is a new phenomenon which results from the implementation of Hilbert space
positivity in terms of stringlocal fields for $s\geq 1$ interactions and has
no counterpart in pointlocal models.

The recent proposal for a Hilbert space based formulation is not the first
attempt to avoid the Krein space setting of $s=1$ gauge theory. Already in
the 60s DeWitt \cite{DeWitt} and Mandelstam \cite{Man} explored the
possibility of circumventing gauge theory by implementing interactions
directly in terms of Hilbert space-compatible field strengths. But without
the awareness of the short distance singularity-reducing role of directional
fluctuations of stringlocal vectorpotentials, a renormalizable theory in
Hilbert space cannot be formulated. More important for the present Hilbert
space based setting was the existence of a powerful structural theorem for
QFTs with compact localizable (pointlike generated) observable subalgebras
and a positive energy representation of the Poincar\'{e} group for which the
energy momentum spectrum contains a mass gap \cite{Bu-Fr}. Whereas the
validity of scattering theory in the presence of a mass gap is hardly
surprising, the assertion that the generating algebras of the full theory
can be localized in (arbitrarily narrow) spacelike cones is somewhat
unexpected.

The previous observations about possible clashes between the pointlike
localization of tensorpotentials and the Hilbert space positivity, as well
as their resolution by working instead with stringloal fields, find their
natural explanation in this theorem. For zero mass vectormesons one cannot
rely on this theorem; in that case the use of the quantum Gauss theorem \cite%
{Bu} leads to the stringlocal nature of Maxwell-charge-carrying fields \cite%
{Bu}. Those charged strings are \textit{rigid}, in particular their
direction cannot be changed by Lorentz transformations (spontaneous
breakdown of Lorentz symmetry \cite{Froe})

The re-translations of these findings from the algebraic "local quantum
physics" (LQP) setting \cite{Haag} into that of (operator algebra-)
generating fields\footnote{%
Whereas the core of (causally closed compact) double cones are points, that
of (causally closed noncompact) spacelike cones are covariant semi-infinite
spacelike linear strings.} (operator-valued distributions) means that in any
QFT with a mass gap, which contains a pointlike generated observable
subalgebra which generates the vacuum sector, the superselected charged
sectors can be described in terms of stringlike covariant fields $\Psi
(x,e).~$In this terminology "pointlike" is the special case of stringlike
namely $e$-independence. To generate QFTs in terms of fields one does in
particular not need fields which are localized on hypersurfaces
("branelike"-fields). The theorem does however not say anything about
whether a particular concrete model can be generated by pointlike fields or
if stringlocal fields are necessary. Here our perturbative results connect
the necessity for using stringlike localization with the breakdown of
renormalizability for their pointlike counterparts. In particular $s\geq 1~$%
interactions have an interaction density with short distance scale dimension 
$d_{int}>4$ require the use of stringlike localization. The connection
between nonrenormalizability and weakening of pointlike localization is,
according to my best knowledge, a new result which will have consequences on
earlier attempts to relate nonrenormalizability of certain models with
modified renormalization group behavior \cite{Reuter}. The interesting
question of an analog of the Callan-Symanzik equations for stringlocal
interactions will remain outside the scope of the present paper. Gauge
theories and the stringlike Hilbert space formulation as well as their
relation to each other is the main topic of this note.

The next section entitled "kinematical prerequisites" presents the
construction of the massive stringlike interaction-free vectorpotential $%
A_{\mu }(x,e)$ in terms of their pointlike Proca siblings $A_{\mu }(x)~$and
a scalar escort field $\phi (x,e)$ which are all members of the same
stringlocal localization class. Whereas the massive Wilson loop in terms of
the stringlocal field is equal to that in terms of the Proca field and hence
its $e$-independence is manifest, its zero mass limit is $e$-independent in
a topologically more subtle way. This explains the breakdown of Haag duality
for multiply connected spacetime regions with interesting relations to the
quantum mechanical Aharonov-Bohm effect; in fact this duality breakdown is a
generic property of all \textit{massless} $s\geq 1$ potentials \cite{B1} 
\cite{B2}. The use of the indefinite metric pointlike potentials leads to
wrong results, in other words the gauge description already breaks down if
one considers global gauge invariant objects as Wilson loops; it remains
strictly limited to pointlike generated gauge invariant local observables.

In the third section the kinematical preparation is used for the calculation
of the string-independent second order S-matrix of massive vectormesons
interacting with matter (massive scalar QED and the coupling to Hermitian $H$%
-fields\footnote{%
The letter $H$ stands for both: Hermitian and Higgs.}). For the $H$-coupling
the "Mexican hat" potential, which has been imposed for the implementation
of the alleged symmetry breaking "Higgs mechanism", emerges instead as a
second order "induced potential" of the renormalizable coupling of a massive
stringlocal vectormeson to \ $H$ fields; \textit{it is not the breaking of
gauge symmetry but rather its upholding} which in massive vectormeson-$H$
couplings induces \textit{a second order Mexican hat potential. }In the new
stringlocal Hilbert space setting this arises from the string-independence
of the second order S-matix.

The concluding remarks present a resum\'{e} as well as an outlook. Here
additional remarks about consequences of the stringlocal $s\geq 1~$%
in~Hilbert space (SLF) can be found and the stringlocal scenario for the
expected confinement in the zero mass limit of self-interacting massive
vectormesons is presented.

\section{Kinematical prerequisites}

With an s=1 pointlike Proca field%
\begin{eqnarray}
&&A_{\mu }^{P}(x)=\frac{1}{(2\pi )^{3/2}}\int e^{ipx}\sum_{s_{3}}u_{\mu
}(p,s_{3})a^{\ast }(p,s_{3})+h.c.  \label{pro} \\
&&\left\langle A_{\mu }^{P}(x)~A_{\nu }^{P}(x^{\prime })\right\rangle =\frac{%
1}{(2\pi )^{3}}\int e^{-ip(x-x^{\prime })}M_{\mu \nu }(p)\frac{d^{3}p}{2p_{0}%
},~~M_{\mu \nu }(p)=-g_{\mu \mu ^{\prime }}+\frac{p_{\mu }p_{\mu ^{\prime }}%
}{m^{2}}  \notag
\end{eqnarray}%
one can connect two stringlocal fields%
\begin{eqnarray}
A_{\mu }(x,e) &=&\int_{0}^{\infty }F_{\mu \nu }(x+se)e^{\nu }ds,~F_{\mu \nu
}(x)=\partial _{\mu }A_{\nu }^{P}(x)-\partial _{\nu }A_{\mu }^{P}(x)
\label{def} \\
\phi (x,e) &=&\int_{0}^{\infty }A_{\mu }^{P}(x+se)e^{\mu }ds  \notag
\end{eqnarray}%
which are linearly related%
\begin{equation}
A_{\mu }(x,e)=A_{\mu }^{P}(x)+\partial _{\mu }\phi (x,e)  \label{proca}
\end{equation}%
This relation may also be directly derived in terms of the $u$-intertwiners
of the three fields\footnote{%
I thank Jens Mund for pointing out that these relations correspond to linear
relation between stringlocal intertwiners which have been introduces in \cite%
{MSY}} defined in (\ref{def}) for the computation of the intertwiners of the
fields defined in (\ref{def}). Since in the algebraic LQP setting fields
which are relatively local (i.e. members of the same stringlocal Borchers
class \cite{St-Wi}) with respect to each other are considered as different
"field-coordinatizations" of the same model (the same physics), this zero
order relation is the kinematical prerequisite for its continued validity
for the interacting potentials (its implementation is part of the
renormalization process). For the perturbative Bogoliubov S-matrix only the
free relation will be needed.

\textit{At this point one begins to understand why the early attempts of
Mandelstam and DeWitt failed. What was missing were two interrelated
properties, namely the short distance scale dimension-reducing directional
fluctuations of stringlocal potentials and the perception that the Hilbert
space positivity requires the presence of a scalar "escort" }$\phi ~$\textit{%
of the vectorpotential. This escort field does not generate new degrees of
freedom; together with its "mother potential" }$A_{\mu }(x,e)~$\textit{it is
simply the result of lowering the }$d_{sd}=2~$\textit{of the Proca potential 
}$A^{P}(x)$\textit{\ in order to obtain an interaction density which is
below the power-counting limit }$d_{int}\leq 4.~$\textit{We will see that
the escort enters the interaction density in an essential way. }

A useful reading of this relation (equivalent to the picture of lowering of $%
d^{P}=2$ to $d^{S}=1$ by one unit of $d$ going into the $e$-fluctuations) is
to say that the derivative of the $d=1$ scalar stringlocal "escort field" $%
\phi ~$compensates the leading short distance singularity at the price of
weakening the localization from point- to stringlike. Each vectormeson
potential has its $\phi $-companion which shares the degrees of freedom and
the mass. It turns out that this mechanism permits a generalization to
arbitrary high integer spin in which case there appear $s$ stringlocal
escort $\phi ~$fields with spins between $0$ and $s-1$; the scalar $\phi ~$%
enters with $s$ derivatives and the tensor indices of the different $\phi $
together with the number of derivative in front always add up to $s$. In all
cases the relation breaks down in the massless limit since there are neither
pointlike Proca fields nor stringlocal massless $\phi ^{\prime }s$ within
the mentioned spin range; in this limit only the stringlocal $A_{\mu }$ or
equivalenty the linear combination (\ref{proca}) survives-

Our main interest is $s=1.$ In this case the integration of the equation
along a closed spacelike circle leads to%
\begin{eqnarray}
\doint A_{\mu }(x,e)dx^{\mu } &=&\doint A_{\mu }^{P}(x)dx^{\mu },~m>0
\label{loop} \\
\doint A_{\mu }(x,e)dx^{\mu } &=&\doint A_{\mu }(x,e^{\prime })dx^{\mu
},~~\forall e,e^{\prime },~m=0  \notag
\end{eqnarray}%
since the integral along a spatial path of a gradient of $\phi ~$is the
difference between the endpoint and the initial point which vanishes for
coinciding points. The resulting independence of the Wilson loop operator
from the direction $e$ corresponds to the expected gauge invariance of the
loop. In the zero mass limit, the difference of two line integral over the$~$%
same curve but with $e^{\prime }s~$pointing into different directions is
simply the difference between the $\phi (x,e)$ and $\phi (x,e^{\prime })$.
Although the individual $\phi $ are infrared divergent, their difference at
the same point (the initial and final point of the Wilson loop) but with
different directions stays infrared finite; hence the difference between two
Wilson loops over stringlocal potentials with different string direction
vanishes (the second line in \ref{loop}). In this case there remains a
topological imprint which the string dependence leaves behind.

The result is the breakdown of Haag duality for multiply connected spacetime
regions. In such a case the algebra of a causally closed region $\mathcal{A(O%
}),$ $\mathcal{O=O}^{\prime \prime }$ is not the same as the commutant of
the algebra $\mathcal{A}(\mathcal{O}^{\prime })$~rather one finds%
\begin{equation}
\mathcal{A(O})\varsubsetneq \mathcal{A(O}^{\prime })^{\prime }  \label{dual}
\end{equation}%
for $\mathcal{O}$ a spacelike torus. The "thickened" Wilson loop is an
operator in the right hand algebra which is not in $\mathcal{A(O}).$ The
physical picture is that there are operators localized in a spacelike
separated intertwining torus which have to penetrate the cylinder subtended
from the loop into the $e$-direction somewhere whatever direction of $e$ one
chooses. The bigger the helicity, the higher is the genus up to which new
violations of Haag duality occur.

This effect can also be directly derived from the equal time commutation
relations of the$~$field strengths without the use of their stringlike
tensorpotentials (for $s=1$ this was shown in the unfortunately unpublished
work \cite{LRT}). The important point here is that the proof in terms of
stringlocal potentials is simpler whereas the use of the Stokes theorem for
pointlike potentials in Krein space is misleading. This is the simplest
illustration for the necessity of a Hilbert space setting for potentials and
the ensuing SLF setting. It should be clear that the stringlike localization
is not the result of a playful fancy of particle theorists who set out to
"try something else" and look for its potential physical use later on (which
led to String Theory \cite{hol}), but rather the consequence of maintaining
the Hilbert space positivity of quantum theory also for $s\geq 1$.

To avoid misunderstandings about the aim of the present work, it should be
said that the central issue of this paper is not the change of the viewpoint
about the nature of the QFT analog of the Aharonov-Bohm effect in the
Hilbert space setting (instead of the usual gauge-theoretic pointlike
setting in Krein space) but rather the changes which Hilbert space
positivity causes in the ongoing research of the Standard model (in
particular the Higgs issue and confinement). The simplicity of the above
presentation of interaction-free spacetime loops is meant as a pedagogic
illustration that even free field properties of \textit{global} gauge
invariants come out incorrect; the physical range of gauge theory is
strictly limited to gauge invariant local observables.

Thinking of the functorial relation between Wigner's positive energy
representation spaces for the Poincar\'{e} group and interaction-free
quantum fields and the associated local nets of operator algebras, it is
interesting to note that the operator algebraic breakdown of Haag duality
for multiply connected spacetime regions has a \textit{spatial counterpart
in modular localized toroidal subspaces of the Wigner representation space}.
The spatial counterpart of the stringlocal vectormeson field is the
covariant stringlocal Wigner wave functions which together with its opposite
frequency part defines a hyperbolically propagating classical wave functions
for a classical stringlocal vectorpotential. Different from the standard use
of pointlike classical vectorpotentials, the stringlocal vectorpotential
extends to spacelike infinity and thus prevents the formation of compact
Cauchy data for potentials. 

The Wilson loops formed with the correct classical potentials, although
being causally separated from a magnetic flux inside the Wilson ring, still
"feel" its presence because the flux lines have to penetrate the walls of
one of the infinite spatial cylinders which are associated to the different
choices of the $e^{\prime }s;$ this is a topological imprint which the
stringlocal vectorpotential leaves behind if one forms a classical Wilson
loop which looses geometric memory of the $e$ of its vectorpotential. This
is the way in which the presence of the magnetic flux is perceived despite
the geometric causal separation between the potential in the loop and the
magnetic flux inside. The Aharonov-Bohm effect is a quasiclassical relic
(quantum mechanical matter in a classical vectorpotential) of this breakdown
of Haag duality and admits higher helicity generalization ($m=0,\left\vert
h\right\vert ~finite$); it looses its exotic appeal by abandoning the idea
of pointlike potentials. The observance of this discrepancy between the
naive geometric picture of apparent spacelike separation and the classical
limit of the more hidden stringlocal nature of quantum potentials removes
all feelings of nonlocal magical aspects of the A-B effect. 

This provides strong support for the Hilbert space formulation of $s\geq 1$
fields as compared to the gauge theoretic setting. It shows that the often
as magic perceived mismatch between the naive geometric view and that coming
from causal localization of the A-B effect disappears if one permits the
more fundamental QFT the chance to tell its classical counterpart to use the
correct description (classical stringlocal fields and their topological
implication for loops). This philosophy is opposite to that of
"quantization" where the (Euler-Lagrange) description is used as a classical
crutch to enter the world of QT. Stringlocal fields in Hilbert space are not
solutions of Euler-Lagrange equations and the consistency of gauge theory
with QT is limited to the vacuum sector (and even there is appears more a
matter of luck than of exigency).

It is also interesting to note that the better known "Coulomb gauge" is a
Hilbert space description which results from the covariant stringlocal
potential by directional $e$averaging within a spatial hyperplane%
\begin{equation}
A_{\mu }^{C}(x)=\int A_{\mu }(x,\vec{e})\frac{d\Omega }{4\pi }%
,~A_{0}^{C}(x)=-\frac{1}{4\pi }\int d^{3}y\frac{\func{div}\vec{E}(\vec{y},t)%
}{\left\vert \vec{y}-\vec{x}\right\vert },~\vec{A}^{C}(x)=\frac{1}{4\pi }%
\int d^{3}y\frac{rot\vec{H}(\vec{y},t)}{\left\vert \vec{y}-\vec{x}%
\right\vert }  \label{coul}
\end{equation}%
This exposes its covariance property in terms of the action of the Lorentz
group on the time-like vector orthogonal to the hypersurface.

In massive $s\geq 1$ theories the clash between Hilbert space and pointlike
localization and its resolution through the use of stringlike
tensorpotentials is reflected in the fact that behind pointlike
nonrenormalizability there looms a weakening of localization; the attempt of
a pointlike description leads to singular matter fields with short distance
dimension $d=\infty $ (unlimited increase with the perturbative order).
Mathematically the formal pointlike fields are singular to such an extend
that a smearing with all compact supported spacetime testfunctions is not
possible. The Wightman localization property can only be recovered in terms
or renormalizable stringlocal fields $\Psi (x,e)$ which can be smeared with
all Schwartz testfunctions $f(x,e)$\ with compact supports in $\mathfrak{D}$(%
$\mathbb{R}^{4}~$tensored with $3$-dim. $de~Sitter^{~}),~$see next section$%
.~ $

Linear relations between high dimensional pointlike fields and their lower
dimensional stringlike siblings (which for s=1 reduce to (\ref{proca})) are
the key for the conversion of nonrenormalizable pointlike interaction
densities into affiliated renormalizable stringlocal interactions. Here are
two illustrations:

\begin{itemize}
\item Scalar massive QED with the first order interaction density (products
of operators are always Wick-ordered products) which is to be multiplied by
a numerical coupling strength%
\begin{eqnarray}
L^{P} &=&A_{\mu }^{P}j^{\mu },~j^{\mu }=i\varphi ^{\ast }\overleftrightarrow{%
\partial ^{\mu }}\varphi  \label{1} \\
L^{P} &=&L-\partial ^{\mu }V_{\mu },~L=A_{\mu }j^{\mu },~V_{\mu }=\phi
j_{\mu }  \notag
\end{eqnarray}%
here we used (\ref{proca}). Whereas the $L^{P}$ has operator short distance
dimension $d=5,$ which is too high in order to be within the renormalizable
power counting range of $d\leq 4,$ the stringlocal action has $d=4.~$The
incriminated dimension 5 has been absorbed into a derivative term where it
can be disposed of in the adiabatic limit; in this way the first order
S-matrix of the pointlike $L^{P}~$is the same as that of the stringlike $L.~$%
In fact one does not have to know the polynomial expression, rather its form
together with that of $V_{\mu }~$results from the requirement that $%
L-\partial V$ is independent of $e.$ We may say that, given the field
content, the renormalizable first order interaction is \textit{self-induced
and (in the cases tested up to now) unique.~}This induction principle
extends to all orders (see next section).

\item The coupling of a massive vectormeson to a Hermitian field is%
\begin{eqnarray}
L^{P} &=&A^{P}\cdot A^{P}H=L-\partial ^{\mu }V_{\mu }  \label{2} \\
L &=&A\cdot (A^{P}H+\phi \partial H)-\frac{m_{H}^{2}}{2}\phi ^{2}H,~~V_{\mu
}=A_{\mu }^{P}\phi H+\frac{1}{2}\phi ^{2}\partial ^{v}H  \notag
\end{eqnarray}%
up to renormalizable $H^{3},H^{4}~$self-interactions whose coupling
strengths turn out to be fixed by higher order induction so that at the end
there is just one first order coupling $g~$and the two masses of the
interaction-defining fields. The latter are usually not counted but for
comparison with the Higgs mechanism it is helpful to mention them.
\end{itemize}

In both cases the leading short distance singularity is "peeled off" from
the pointlike scalar $L^{P}$ in terms of a divergence of a vector in analogy
to the peeling off in (\ref{proca}) by the gradient of a scalar. The $d=5$
of the pointlike interaction density has been converted into the
power-counting-compatible $d=4~$of the stringlike density. The split into a
stringlocal $d=4\ $interaction and a divergence is unique up to
divergence-free additional contributions to $V_{\mu }.$

\section{Higher order string independence of the S-matrix and induced
Mexican hat potentials}

If the construction of higher order interaction densities $%
TL(x_{1},e_{1})...L(x_{n},e_{n})$ would not involve the time-ordering of
operator-valued distributions, the derivatives in $\partial ^{\mu }V_{\mu }$
could be taken outside the time-ordering and the previous relations (\ref{1}%
\ref{2}) would have a straightforward $n^{th}$ order extension. The
singularities at point- and string-crossings prevent this, and in case of
interactions between pointlike fields Epstein and Glaser \cite{Ep}
established rules for the inductive construction of time ordered products
based on the perturbative implementation of causality with minimal scaling
degree (which is not larger than the naive high energy divergence degree of
corresponding Feynman integrals). Those couplings for which the minimal
scaling degree stay finite independent of the perturbative order are called
renormalizable. They are those couplings for which the scaling degrees
remains finite and the pointlike fields remain localizable. A simple
criterion for renormalizability is $d_{L}\leq 4$ where $d_{L~}$is the short
distance scaling degree of $L;~$the fulfillment of this power-counting
criterion the prerequisite for renormalizability. What renders
nonrenormalizable theories physically worthless (apart from possible
phenomenological use) is not so much the unbounded increase for $%
p\rightarrow \infty $ but primarily the growth of the number of coupling
parameters associated with an ever increasing number of counterterms.

An systematic extension of the E-G method to string-crossings has not yet
been published \cite{Jens1}, but fortunately such systematics is not yet
needed if, as in the present work, one's aim is to direct attention to the 
\textit{derivation of new concepts and results} from explicit second order
model calculations. The main new message is that requirement of string
independence of the S-matrix places new restrictions on renormalizations
which lead to the concept of \textit{induced normalization terms} i.e. their
couplings are uniquely determined in terms of the basic model-defining first
order coupling and the masses and spins of the interaction-defining free
fields. This is similar to the gauge theoretic setting where the second
order $A_{\mu }A^{\mu }~$term of scalar QED is induced from the first order
coupling by gauge symmetry. The fundamental difference is however that in
the new setting one does not have to invoke a new symmetry, rather the
induction is the consequence of causal localization in a Hilbert space
setting (modular localization). The main point here is not that the induced
counterterm structure is more or less isomorphic\footnote{%
Most interesting are however additional terms in the SLF setting which have
no counterpart in the BST gauge formalism.} to that of the gauge setting,
but rather that the Hilbert space positivity has additional physical
consequences about the structure of global string-independent quantities
which the gauge setting cannot reproduce; the simplest free field
illustration is the loss of the above topological property in the used of
Wilson loops.

In second order the $e$-independence is more subtle since the direct
definition of the pointlike $TL(x)^{P}L^{P}(x^{\prime })$ fails as a result
of nonrenormalizability. However the requirement%
\begin{equation}
d_{e}(TLL^{\prime }-\partial ^{\mu }TV_{\mu }L^{\prime })=0,  \label{indep1}
\end{equation}%
where the bracket is considered as a zero differential form in $e$ (on the
3-dim. unit de Sitter space) on which a differential form operator $d_{e}$
acts, is perfectly meaningful since the short distance degrees of $LL$'~and~$%
V_{\mu }L^{\prime }$ does not exceed 8. A more symmetric form which
simultaneously takes care of $e,e^{\prime }$ is%
\begin{eqnarray}
&&d(TLL^{\prime }-\partial ^{\mu }TV_{\mu }L^{\prime }~-\partial ^{\mu
\prime }TLV_{\mu }^{\prime }+\partial ^{\mu }\partial ^{\nu \prime T}V_{\mu
}V_{\nu }^{\prime })=0,~~d=d_{e}+d_{e^{\prime }}  \label{indep2} \\
&&T(LL^{\prime })^{P}:=TLL^{\prime }-\partial ^{\mu }V_{\mu }L^{\prime
}~-\partial ^{\mu \prime }LV_{\mu }^{\prime }+\partial ^{\mu }\partial ^{\nu
\prime }V_{\mu }V_{\nu }^{\prime }  \notag
\end{eqnarray}%
Although a direct perturbative treatment of the pointlike second order
interaction density would conflict with renormalizability, the indirect
definition in terms of the string-independent bracket as in the second line
is perfectly reasonable. It is again the encoding of the highest scale
dimensions into derivatives of lower dimensional operators within the power
counting limit ($d=4n$ in $n^{th}$ order) which permits to define pointlike
products whose direct $n^{th}$ order scale dimension $d=4n+n$ would exceed
the power-counting limit and, if treated in the usual pointlike setting,
would lead to an ever increasing number of coupling parameters since the
induction mechanism would be absent. The relations (\ref{indep1},\ref{indep2}%
) have straightforward $n^{th}$ order extensions, but here their
perturbative model implementation will be limited \ to second order.

The similarity of (\ref{proca}) with a gauge transformation suggest that the
connection of the stringlocal matter field and its strongly singular (not
Wightman-like, pointlike nonrenormalizable) pointlike sibling should be 
\begin{equation}
\psi (x)=e^{-ig\phi (x,e)}\psi (x,e)
\end{equation}%
However the conceptual content of these relations is quite different than
that of gauge transformations; they do not stand for a symmetry
transformation ("gauge symmetry") but rather relate two
"field-coordinatizations" in Hilbert space which belong to the same
localization class (relative locality with respect to each other). The proof
of these off-shell relations in the presence of interactions requires an
extension of the St\"{u}ckelberg-Bogoliubov-Epstein-Glaser (SBEG) on-shell
formalism. In fact the representation for the S-matrix is a special case of
that for fields and their correlation functions; all these formulas require
to take the adiabatic limit\footnote{%
The SBEG field construction is opposite to that of LSZ scattering theory; in
the latter case on starts from the field correlations and computes the
S-matrix as the "cream on the cake".}.

Fortunately in massive QED this is not needed since the matter field only
enters the interaction in form of the conserved current and the SBEG
formalism for the on-shell S-matrix only deals with time-ordered products of
free. The expected off-shell connections between renormalizable stringlike
fields and their formally nonrenormalizable pointlike siblings contains
however the interesting message that the kind of nonrenormalizability
addressed in the present work is a result of the clash of enforced pointlike
localization with the Hilbert space positivity; a clash which can be removed
by passing to the renormalizable stringlocal formulation.

The application of (\ref{indep1}) to the model of massive scalar QED (\ref{1}%
) requires to expand the time ordered product into Wick-products. The
component without contractions fulfills this relation trivially as a
consequence of $d_{e}(L-\partial ^{\mu }V_{\mu })=d_{e}L^{P}=0.$ For the
1-contaction component this is not the case since the anomaly $\mathfrak{A}$ 
\begin{eqnarray}
A &:&=d_{e}(T_{0}LL^{\prime }-\partial ^{\mu }T_{0}V_{\mu }L^{\prime
})_{1-contr.}=-d_{e}N+\partial ^{\mu }N_{\mu } \\
A^{\prime } &=&A(x,e\longleftrightarrow x^{\prime },e^{\prime }),~~\mathfrak{%
A}=A+A^{\prime }  \notag
\end{eqnarray}%
where $T_{0}~$refers~to the "kinematical" time ordering for which all
derivatives act outside the $T_{0~}~$e.g.%
\begin{eqnarray}
\left\langle T_{0}\partial _{\mu }\varphi (x)\partial _{v}^{\prime }\varphi
^{\ast }(x^{\prime })\right\rangle  &=&\partial _{\mu }\partial _{v}^{\prime
}\left\langle T_{0}\varphi (x)\varphi ^{\ast }(x^{\prime })\right\rangle  
\notag \\
\left\langle T\partial _{\mu }\varphi (x)\partial _{v}^{\prime }\varphi
^{\ast }(x^{\prime })\right\rangle  &:&=\left\langle T_{0}\partial _{\mu
}\varphi (x)\partial _{v}^{\prime }\varphi ^{\ast }(x^{\prime
})\right\rangle +ig_{\mu \nu }c\delta (x-x^{\prime })  \label{4}
\end{eqnarray}%
According to the Epstein-Glaser normalization rules the scaling degree 4
(logarithmically diverging) propagator of derivative of scalar fields permit
a scaling degree 4 preserving renormalization in terms of a delta term with
a yet undetermined parameter $c$. The contributions to the anomaly coming
from the action of the divergence $\partial _{\mu }~$on contractions in $%
T_{0}j_{\mu }(x)j_{\nu }(x^{\prime })|_{1-contr}.$are$~$%
\begin{equation}
\partial ^{\mu }\left\langle T_{0}\partial _{\mu }\varphi (x)\varphi ^{\ast
}(x^{\prime })\right\rangle =(\partial ^{\mu }\partial _{\mu
}+m^{2})\left\langle T_{0}\varphi (x)\varphi ^{\ast }(x^{\prime
})\right\rangle -m^{2}\left\langle T_{0}\varphi (x)\varphi ^{\ast
}(x^{\prime })\right\rangle =-i\delta (x-x^{\prime })-reg.
\end{equation}

It is precisely these numerical anomalies of time-ordered propagators which
determine the above operator anomalies. The result of the contraction
combinatorics is 
\begin{equation}
N=\varphi ^{\ast }\varphi A\cdot A^{\prime },~N_{\mu }=\delta \varphi ^{\ast
}\varphi \phi A_{\mu }^{\prime }
\end{equation}%
where the $A_{\mu }^{\prime }$ stands for $A_{\mu }(x,e^{\prime })$ and $%
\delta $ is $\delta (x-x^{\prime }).~$We also use the fact that $%
d_{e}\partial _{\mu }\phi =d_{e}A_{\mu }.~$For the renormalization of $%
T_{0}LL^{\prime }$ the $N_{\mu }~$(which corresponds to the renormalization
of $T_{0}V_{\mu }L^{\prime }$) is not important. Hence the relevant $N~$part
of the full anomaly is symmetric in $e$ and $e^{\prime }$%
\begin{equation}
\varphi ^{\ast }\varphi (A\cdot A^{\prime }+A^{\prime }\cdot A)
\end{equation}%
The reader immediately recognizes that this induced contact term which must
be added to $T_{0}LL^{\prime }$ corresponds to the quadratic term in the
gauge theoretic formulation of scalar QED. In the present setting it is
simply the consequence of the string-independence of the S-matrix. The
philosophy underlying the present setting suggests to absorb this term into
a re-definition (renormalization) of the time ordering by 
\begin{eqnarray*}
T &=&T_{0}~for~propagators~of\text{ }scaling~degree\text{ }d<4 \\
&&T\text{ }as~in~(\ref{4})\text{ }with~c=-1~for\text{ }d=4
\end{eqnarray*}%
In this case the $T_{0}~$contraction of the propagator of the derivative of
the fields coming from the anomaly an be absorbed in the redefinition (\ref%
{4}) $T_{0}\rightarrow T$ of the contraction in $T_{0}LL^{\prime }$ so that
the $e$-independence (\ref{indep1}) is fulfilled with the $T$ ordering
instead of the $T_{0}.$ One can then directly check that (\ref{indep2})
really defines a second order pointlike interaction density without having
to introduce a counterterm with a new coupling (as it would be necessary in
in the pointlike formalism in Krein space before imposing the gauge
invariance condition). In fact the formula (\ref{indep2}) guaranties the
independence of the scattering amplitude on the string directions.

The string-independence in the Hilbert space setting $d_{e}S=0~$corresponds
to $sS=0~$with a nilpotent $s$ in the BRST gauge setting, where the
definition of $s~$requires to enlarge the already unphysical Krein space in
terms of ghost operators. The computation is analogous, except that there
are no$~A_{\mu }^{\prime }.~$The Hilbert space positivity for massive QED
leads to on-shell results for $2\rightarrow 2~$scattering which have the
same formal appearance as those coming from gauge theory, even though the
concepts and calculations are different. In the following we sketch the
operator gauge derivation a la Scharf; this should also be seen as a
recompensation of the early results of the University of Z\"{u}rich group 
\cite{Aste} which were left on the wayside by the Standard Model caravan and
eventually succumbed to the maelstrom of time\footnote{%
We (Jens Mund and myself) only became aware of this contribution after our
work on stringlocal fields \cite{MSY} \cite{Rio}. Here these results have
been rewritten into our formalism in order to facilitate comparisons.}.

The self-induced $e$-independent $d_{e}(L-\partial V)=0~L,V$ pair of a
trilinear $A-H$ coupling (\ref{2}) has its BRST gauge-theoretic counterpart
in $s(L^{K}-\partial V^{K})=0$ where the fields now act in the Wigner-Fock
Krein space. Apart from the fact that in addition to $u^{K}=s\phi ^{K}~$we
also have an anti-ghost field\footnote{%
The ghost field corresponds to $u:=d_{e}\phi ,$ but the anti-ghost has no
counterpart since $s\tilde{u}=(\partial A^{K}+m^{2}\phi ^{K})$ and the
corresponding SLF expression vanishes since it is a an operator relation in
Hilbert space.} $\tilde{u}$\ and the first order induction of a trilinear
interaction leads to (we surprise the superscript $K$ for the individual
fields) and differences in the normalization of the $\phi ^{K}$ field, the
formal expression for the first order gauge interaction is analog to (\ref{2}%
)

\begin{align}
L^{K}& =m\left( A\cdot AH-H\overleftrightarrow{\partial }\phi \cdot A-\frac{%
m_{H}^{2}}{2m^{2}}H\phi ^{2}+bH^{3}+u\tilde{u}H\right) \\
Q_{\mu }^{K}& =m(uA_{\mu }H-\frac{1}{m}u\phi \overleftrightarrow{\partial }%
_{\mu }H)=sV_{\mu }^{K}  \notag \\
sA& =\partial u,~sH=0,~s\phi =u,~s\tilde{u}=-(\partial A+m^{2}\phi )\simeq 0
\end{align}%
Up to the $H$ self-interactions these trilinear terms are the unique (up to $%
V_{\mu }~$contributions with vanishing divergence) renormalizable induced
first order terms. The third line is the action of $s$ on the individual
free fields which act in the Krein space analog of the Fock space. The
second order anomalies $sT_{0}LL^{\prime }-\partial ^{\mu }T_{0}Q_{\mu
}L^{\prime }~$and the corresponding term with $Q\rightarrow L,~L^{\prime
}\rightarrow Q^{\prime }~$(prime denotes $x\rightarrow x^{\prime }$) lead to
the additional (induced) contribution \cite{Scharf}

\begin{eqnarray}
&&T_{0}L^{K}L^{K\prime }+i\delta (x-x^{\prime })(A\cdot AH^{2}+A\cdot A\phi
^{2})-i\delta (x-x^{\prime })R_{scharf}  \label{second} \\
&&R_{Scharf}=-\frac{m_{H}^{2}}{2m^{2}}(\phi
^{2}+H^{2})^{2},~V_{Scharf}\equiv g^{2}R+first~order~H,\phi -terms  \notag \\
&&V_{Scharf}=g^{2}\frac{m_{H}^{2}}{8m^{2}}(H^{2}+\phi ^{2}+\frac{2m}{g}\phi
)^{2}-\frac{m_{H}^{2}}{2}H^{2}  \label{pot}
\end{eqnarray}%
The remaining step consists in absorbing the quadratic in $A$ contributions
into a redefinition of the time-ordered product $T_{0}\rightarrow T~~$(as in
the previous case of scalar massive QED) and to verify that the absence of
third order anomalies requires to introduce a $cH^{4}$ self-coupling and
fixes the parameters $b$ and $c.~$

The net result\ is the $H$-$\phi ~$local potential $R$ of degree 4~(\ref{pot}%
). The appearance of $g^{-1}~$terms results from writing the potential into
the symmetry-breaking form of the Higgs mechanism from where one may read
off the Mexican hat parametrization of the Higgs mechanism (the quartic
self-coupling of scalar QED and the shift in field space). It shows that
that the latter is incompatible with the logic of renormalized perturbation
theory, whereas the mechanism of induction potential is its logical second
order result. The lesson to be learned is that, whereas at the frontiers of
the foundational and sometimes highly speculative research as particle
theory one sometimes is led to use metaphors as placeholders for
incompletely understood issues, the problem starts when the placeholder
nature is not recognized in due time. Admittedly this was not easy since the
coupling of Hermitian fields to massive vectormesons which vanish in the
massless limit is somewhat unaccustomed for physicists coming from QED.     

For the details of the anomaly calculations and the derivation of the (up to
second order) induced potential (\ref{pol}) we refer to Scharf (formula
4.1.38 \cite{Scharf}). As in the case of the Schwinger-Swieca screening of
the Maxwell current of interacting massive vectormesons, we recommend to the
readers to take their time to look up the cited historical papers in order
to convince themselves how close some individuals came to the correct
understanding of interactions involving massive vectormesons.

This should have caused the ringing of bells, but the Standard Model Theory
community was already in the grip of Big Science; it was too late for a
perception and a critical discussion of these findings; more recent attempts
to remind the particle theory community of some of these forgotten (or never
noticed) results remained without avail \cite{Garcia}. What was still
missing in order to realize that the unfortunate broken symmetry picture is
inconsistent with the computed result was to notice that the
Schwinger-Swieca screening \cite{Sw} of the Maxwell charge of interacting
massive vectormesons (the only conserved current in the $H$-model) is
completely different from the charge associated to the conserved current of
a spontaneously broken symmetry. A deeper sociological analysis of these
strange occurrences in the heart of particle theory will be left to
historians and archeologists of science.

The new SLF Hilbert space formalism does not only confirm these
pre-electronic insights\footnote{%
Setting $e=e^{\prime }~$in the scattering amplitudes, the results formally
agree (up to differences in the $\phi ^{K}~$versus $\phi $ normalizations).}
but also adds an interesting twist to it; the Hilbert space positivity leads
to the presence of additional induced contributions to the Mexican hat
potentials which however vanish on the diagonal $e=e^{\prime }.~$The
definition of pointlike interaction densities \ref{indep2} within the
Hilbert space setting also reveals a new property which has no counterpart
in the gauge setting. These pointlike densities are special cases of the
polynomial expression which one would obtain in second order direct
pointlike nonrenormalizable expression, except that instead of new
counterterm coupling parameters the latter are now not independent but
rather fixed in terms of the first order couplings and masses. As a
consequence the on-shell polynomial degree is lower than its off-shell
counterpart; with other words the scattering amplitudes have a much better
high-energy behavior than that of pointlike correlation functions. This is
the momentum space counterpart of "peeling off" high degree derivative terms
and dispose them in the adiabatic limit, a mechanism which is difficult to
keep track of in momentum space and which shows that the consequences of the
Hilbert space positivity for $s\geq 1~$(which requires the use of stringlike
localization)$~$ cannot be encoded into Feynman graphs.

These deviations of the Hilbert space setting from gauge theory are
interesting, but their detailed derivation go beyond the task set for the
present paper and will be addressed in separate work \cite{M-S1}. As
mentioned in the introduction, the physical credibility of the gauge setting
is restricted to the gauge invariant vacuum sector of local observables
which strictly speaking does not include the global S-matrix. The frequent
formal similarity between the gauge setting and the Hilbert space
formulation may be interpreted as an unexpected extension of the physical
limitations of gauge theory.

In passing it is interesting to see that the screening of the identically
conserved Maxwell current can already be seen in zero order of the Maxwell
current 
\begin{eqnarray}
j_{\mu }^{M} &:&=\partial ^{\nu }F_{\mu \nu },~~~F_{\mu \nu }=\partial _{\mu
}A_{\nu }^{P}-\partial _{\nu }A_{\mu }^{P}=\partial _{\mu }A_{\nu }-\partial
_{\nu }A_{\mu } \\
&&in~zero\text{ }order~~~\partial ^{\nu }F_{\mu \nu }\simeq m^{2}A_{\nu
},~\partial ^{\mu }A_{\mu }^{P}=0  \notag
\end{eqnarray}%
The corresponding zero order Maxwell charge vanishes 
\begin{eqnarray}
&&observation:~Q^{M}=\int j_{0}^{M}(x)d^{3}x,~in~zero\text{ }%
order~j_{0}^{M}(x)\simeq A_{\mu }^{P}(x),~\int A_{0}^{P}(x)=0 \\
&&Thm:The~Maxwell~charge~of\text{ }a~massive~vectormeson~is\text{ }always%
\text{ }screened  \notag
\end{eqnarray}%
Whereas for couplings to Hermitian fields this is the is only conserved
current, the complex matter fields of massive QED also admit another
conserved current whose conserved charge counts the number of charges minus
anti-charges. The charge screening is lost and both currents coalesce in the
massless limit.

It was Schwinger \cite{Schwinger} who conjectured this property of massive
vectormesons which was later proven as a structural (nonperturbative)
theorem by Swieca \cite{Sw} \cite{Swieca}. In this context it may be
interesting to mention that the conserved current in the case of a
spontaneously broken symmetry leads to a divergent charge whose large
distance divergence is due to the presence of a zero mass Goldstone boson.
Interestingly the structural proof of the Goldstone theorem (which has no
relation with nonvanishing one-point functions of fields\footnote{%
The shift in field space is a mnemonic trick to produce a first order
interation with such a divergent charge associated to a conserved Goldstore
current, but it has no intrinsic physical meaning.}) is also due to Swieca 
\cite{E-S} \cite{Car}, in fact he emphasized that the large-distance
diverging charge is the definition of the meaning of spontaneous symmetry
breaking. He probably had the most profound knowledge of both phenomena and
he tried to attract attention away from the incorrect Higgs symmetry
breaking by using in his publication the terminology "Schwinger-Higgs
screening" in his publications. But unfortunately he was unable to stem the
growing tide in favor of symmetry breaking. This and the 40 year reign of
the Higgs spontaneous breaking mechanism instead says a lot about the
present state of post Standard Model.

In retrospect it appears very conspicuous that the same arguments and
equations which started from massless scalar QED and postulated a symmetry
breaking (gauge symmetry?) by performing a "field shift" of a gauge variant
field were almost simultaneously independently presented by at least 3 other
authors/groups of authors in addition to Peter Higgs. For fairness it should
be mentioned that during the first years after these proposals there were
also publications which pointed out that gauge symmetry is not a physical
symmetry but rather a method to extract local observables from a Krein space
setting and as such it is not fit for being broken. Note that the above
induced Mexican hat potential can also be obtained by imposing BRST gauge
invariance on a massive vectormeson-$H$ interactions \cite{Scharf}; Hence it
is it imposition of gauge invariance in interactions of massive $A_{\mu }$
with Hermitian fields and not its breaking in massless scalar QED which is
consistent with the foundational principles of QFT.

These remarks should not be understood as coming from someone whose research
was motivated by the desire to revolutionize particle theory. To the
contrary it is the increasing worry that a particle theory which lost its
contact with its own pre-electronic past has moved into a blind allay which
motivated the author to write this article. An important observation in the
aftermath of the discovery of the Standard Model as the
"Schwinger-Higgs-Swieca screening mechanism" of massive vectormesons should
never have been allowed to be lost in the maelstrom of time. More conceptual
and historical observations which receive their support from the ongoing
impact of the new SLF Hilbert space setting can be found in \cite{hol}. The
many updates of this article since 2011 reflect the continuous development
of these ideas.

The biggest gain of insight from the new $s\geq 1$ Hilbert space setting\ is
expected in the area of self-interactions between massive vectormeson (Y-M
couplings). One expects of a setting whose principle task is to classify
renormalizable interactions within the Hilbert space setting of
renormalizable $d_{e}~$-induced first order interaction densities that it
achieves something equivalent to what in classical gauge field theory is
obtained with the help of the mathematical fibre-bundle setting. For
self-interacting massive vectormesons of \textit{equal masses} the couplings
should be restricted to just one coupling strength. In particular for three
mutually interacting massive gluons of equal mass the result should look like

\begin{eqnarray}
&&L=\sum \varepsilon _{abc}(F^{a,\mu \nu }B_{\mu }^{b}A_{\nu
}^{c}+m^{2}B^{a,\mu }A_{\mu }^{b}\phi ^{c}),~V^{\mu }=\sum \varepsilon
_{abc}F^{a,\nu \mu }(A_{\mu }^{b}+B_{\nu }^{b})\phi ^{c}  \label{ren} \\
&&d_{e}(L-\partial ^{\mu }V_{\mu })=0.~Ansatz~L=sum~of~4~terms~in~A,\phi 
\notag \\
&&\sum f_{abc}^{1}F^{a,\mu \nu }A_{\mu }^{b}A_{\nu }^{c},~\sum
f_{abc}^{2}A^{a,\mu }A_{\mu }^{b}\phi ^{c},~\sum f_{abc}^{3}A^{a,\mu
}\partial _{\mu }\phi ^{b}\phi ^{c},~\sum f_{abc}^{4}\phi ^{a}\phi ^{b}\phi
^{c}
\end{eqnarray}%
where for notational convenience we used the notation $B=A^{P~}$for the
pointlike Proca potentials. The last line denotes the 4 structures which can
contribute to $L.~$The requirement that $L-\partial V~$is $e$-independent
(pointlike) is very restrictive and leads to the first line where the $%
V_{\mu }~$is determined up to a term with vanishing divergence\footnote{%
Similar conclusions follow from the imposition of $s$-invariance of the
S-matrix \cite{Scharf}..}. If one defines the pointlike $L^{P}$ as the
content of the bracket than the first order pointlike S-matrix is identical
to its stringlike counterpart or equivalently: the two $L^{\prime }s~$are
adiabatically equivalent (the boundary term from the divergence of $V_{\mu }$
vanishes in the adiabatic limit) 
\begin{equation}
\int L^{P}=\int L,~~~L^{P}\overset{AE}{\simeq }L
\end{equation}%
This is the beginning of an extremely restrictive \textit{induction mechanism%
} which has no counterpart in the nonrenormalizable pointlike $s\geq 1$
setting. For the full Lie-algebra structure one has to proceed to the
induced second order \cite{M-S2}.\ 

These observations generalize those which were already made in the abelian
case namely the locality principle \ together with Hilbert space positivity
leads to restrictions between couplings which correspond to those of
classical gauge theory (the geometry of fibre bundles). Here they are simply
the result of the Hilbert space positivity which for interactions which
couple $s\geq 1$ fields requires the use of string-localization. There is
absolutely no need for any support from the fibre-bundle setting of
classical gauge theory; QFT does not need any "crutches" from classical
field theory such as those provided by the classical-quantal parallelism of
quantization. Any quantum fields obtained from covariantizing Wigner's
classification of positive energy representation of $\mathcal{P}$ can be
coupled to a scalar density which defines the first order interaction
density of a QFT and in case its short distance dimension falls within the
power-counting range $d_{sd}\leq 4~$the interaction density is on the best
way to define a renormalizable model of QFT. The above "self-induction"
mechanism also works for unequal masses; in this case the $f^{\prime }s$
depend also on mass-ratios.

The full calculation up to second order will be contained in \cite{M-S2}.
Our main point here is that the specification of the prescribed \textit{%
field content together with the }$d_{e}$\textit{\ induction requirement and
the renormalizability restriction }$d_{int}\leq 4~$\textit{from
power-counting determines the interaction }$L$\textit{\ and the (in the
S-matrix adiabatic limit disappearing) divergence contribution of }$V_{\mu }$%
\textit{\ }(again up to $\partial \tilde{V}=0$ terms).\textit{\ Possibly
missing restrictions are expected to come from second order. }This means in
particular that one does not have the freedom of classical field theory to
add interactions; the field content, the power-counting restriction together
with the induction mechanism fix the form of the $s\geq 1$ quantum
interaction. In particular the second order contributions required by
classical gauge invariance cannot be imposed; rather the induction property,
behind which hides the powerful Hilbert space positivity, governs the form
of the quantum interaction.

This is particularly interesting for $s=2$ interactions where the massive
stringlocal potentials $g_{\mu \nu }(x,e)$\textit{\ }associated to their
pointlike "Proca" siblings $g_{\mu \nu }^{P}(x)~$and two escort fields $\phi 
$ and $\phi _{\mu ~\ }$already enter the lowest order interaction density;
this rigidity of the induction remains in the massless limit and stands in
an interesting contrast with the Lagrangian quantization of the classical.
Einstein-Hilbert action from which !quantum gravity" could profit. In this
connection it is important to point out that renormalization group
properties (analogs of the Callen-Symanzik equations) can only be expected
to hold for the stringlocal Wightman fields (and not for the singular
pointlike fields). Even if the asymptotic freedom property would not have
been based on a consistency property but rather on the beta function of an
established Callan-Symanzik equation of massive (infrared-finite) QCD in the
pointlike gauge setting, its physical relevance would still leave doubts.
Only a derivation based on the SLF Hilbert space setting can definitely
close this issue.

The formal $m\rightarrow 0$ limit in the stringlocal interaction density is
generally not sufficient for understanding the content of the massless
limit. If there are off-shell (logarithmic) infrared divergences this should
be taken as an indication of a radical change in the field-particle
relation. Confinement manifests itself in the sense of vanishing of
correlations for $m\rightarrow 0~$which contain in addition to pointlike
composite also stringlike gluons or quarks. This is expected to be obtained
by resummation techniques of the leading $m\rightarrow 0$ logarithmically
divergent terms \cite{hol}.

\section{Resum\'{e} and outlook}

The Hilbert space positivity for higher spin interactions leads to a
weakening of localization and requires the replacement of
nonrenormalizability of pointlike fields by stringlocal interaction
densities. This replaces the BRST gauge theory in Krein space whose physical
range of validity of gauge theory is limited to (pointlike generated) gauge
invariant local observables; global gauge invariant operators as Wilson
loops are already outside its range since they miss the topological origin
of the breakdown of Haag Duality (for $s=1$ the Aharonov-Bohm effect).. In
the SLF Hilbert space setting fields are generically stringlocal and
physical; the subset of pointlike fields correspond to the gauge invariant
local observables of gauge theory.

The SLF formalism leads to profound conceptual and computational changes.
There are 3 basic classes of renormalizable couplings of massive
vectormesons: couplings with complex matter, with Hermitian $H~$matter and
couplings among themselves including a combination of these 3 classes of
couplings; in all cases the masses of vectormesons and the matter fields (in
accordance with the principles of QFT) belong to the the model defining
field content in terms of which the first order interaction density is
defined. The perturbative renormalization formalism secures that these input
masses remain equal to the masses of the elementary particles of the model.
Up to this point there is no difference to the $s<1$ pointlike
renormalization theory. What is however new for $s\geq 1$ is that in a
Hilbert space setting renormalizability requires the vectormeson fields to
be stringlocal and escorted by fields $\phi $, a property which through
higher order perturbation also "infects" the matter fields. As a consequence
the demand that certain physical global objects as the S-matrix stay
string-independent ($e-$independent) leads to a new phenomenon called
induction; instead of counterterms with free coupling parameters the
coupling strengths of the induced terms are determined in terms of the
interaction-defining first order coupling including the masses of the
coupled fields. For the $H-$coupling the induced potential in $H$ and the
escort field $\phi $ has the form of a Mexican hat. In the operator gauge
setting\footnote{%
In the more common functional setting of perturbation theory (Feynman
graphs) the induction mechanism is easily overlooked.} the result is (up to
difference in normalization) the same but the SLF derivation of the Mexican
potential from the BRST $s-$invariance of the S-matrix takes place outside
the narrow limitation of gauge-invariant local operators and hence has less
physical credibility.

The result vindicates the Schwinger-Swieca charge-screening mechanism behind
the $H$-coupling of the Higgs model. Since Schwinger's screening conjecture
referred to massive QED, the coupling of a Hermitian field which first
appeared in the somewhat metaphoric veil of the Higgs model was by all means
an enrichment of possibilities of couplings of massive vectormesons with
matter. Apart from the omnipresent stringlocal scalar Hermitian escorts $%
\phi ,~$all couplings of massive vectormesons to matter (including the $H$%
-coupling) can be removed by applying Occam's razor and therefore are hardly
of the foundational significance which the mass-creating Higgs mechanism
attributes to them. The appearance of a massive gluonium state with the
quantum numbers attributed to the Higgs particle in a system of
self-interacting massive vectormesons would change the conceptual situation.
In such cases of such bound state problems (e.g. hadrons in terms of quarks)
one usually looks for phenomenological description and the $H$-coupling may
well be such a description.

Significant progress from the new ideas is expected in the hitherto unsolved
problems which hide behind the perturbative logarithmic $m\rightarrow 0$
infrared divergencies of stringlocal fields. Here the Hilbert space
positivity is expected to show its full strength; the induction mechanism
for the lowest order self-interactions between massive vectormeson relates
the possible as independent presumed couplings; for equal masses the
trilinear first order couplings must be antisymmetric and the second order
induction is expected to have a Lie-algebra structure. In other words the
properties which hitherto entered via the classical geometry of
fibre-bundles and quantization into the QFT of $s=1$ interactions (and
finally took the form of operator BRST gauge theory) can obtained without
classical crutches alone from the Hilbert space positivity of the SLF
formulation of interactions between massive vectormesons.

There is another important attention-demanding aspect of the Hilbert space
based perturbative stringlocal formulation of $s\geq 1$ interactions.
According to the LSZ reduction formula the scattering amplitudes are simply
the on-shell restrictions of the momentum space correlation functions. Hence
unless a special mechanism intervenes, one expects that the off-shell high
energy degree is passed to the scattering amplitudes. The "peeling
property", i.e. the disposal of the leading short distance singularities of
the form of divergencies $\partial V$ in the adiabatic limit is easily
overlooked in momentum space, in particular since they cannot be encoded
into Feynman graphs and already occur for tree-contributions to the
scattering amplitude. The validity of graphical representations (Feynman
graphs) is limited to pointlike renormalizable interactions for $s<1;~$%
formally they also hold for the gauge setting in Krein space except that
scattering amplitudes are already outside the range of gauge invariant local
observables. This raises the important question whether the more popular
functional formalism (instead of an operator formulation) remains reliable
for $s\geq 1$ interactions.

Looking back at the history of renormalized perturbation theory it is
conspicuous that there has been no noteworthy conceptual investment with
computational consequences since the post wwII discovery of renormalized
perturbation theory in the wake of Lagrangian quantization. Even the later
extensions of these techniques to nonabelian gauge theories did not require
significant new conceptual investments; as before, prescriptions for the
removal of infinities and consistency checks for the so-obtained finite
expressions was all what was needed. Later conceptual progress which allowed
to derive these results from the iterative perturbative implementation of
the foundational causal localization principle was hardly noticed by the
majority of practicing particle theorists. The "think as (or often after)
you pull up your sleeves and compute along" attitude became the motto of
"robust" and very successful particle theory to the extend that it led to
extreme claims e.g. that any physically inspired mathematically correct
calculation will find its material realization in one of the imagined
parallel universes.

It seems that this successful period in particle theory had reached its
apogee at the time of the discovery of the Higgs mechanism. Identical
Lagrangian manipulations which led to identical conclusions were obtained in
at least four independent publications at approximately the same time (a
rather unique event in the history of particle theory), and while on the
experimental side there has been a steady progress mostly confirming
theoretical ideas, particle theory went through a already 40 year lasting
period of impressive conceptual stagnation. The concepts and even their
formulation (including the mass-giving "fattening" of photons) have remained
identical, and completely reasonable ideas as Schwinger's charge screening
for massive vectormesons and its mathematical derivation from first
principles were allowed to vanish in the maelstrom of time.

This is the clearest indication that the conceptual reservoir of the post
wwII particle theory has been used up and new investments are urgently
needed. The present proposal of extending the renormalizable Hilbert space
operator setting of perturbation theory to higher spins $s\geq 1,~$and
confront the localization problems caused by maintaining the Hilbert space
positivity head-on, is meant as a contribution to this goal.

Finally some remarks about the history behind the present ideas may be
helpful. Although it may be seen as a new startup of old ideas of DeWitt 
\cite{DeWitt} and Mandelstam \cite{Man}, this is not the way in which it
arose. The booster was rather our solution of the old problem of the field
theoretic content of Wigner's infinite spin positive energy representation
of the Poincar\'{e} group \cite{MSY}. Although it was already clear by 1970
that this representation does not admit an associated pointlike field
description \cite{Y}, the first clue in what direction to look came from the
application of \textit{modular localization} to the positive energy Wigner
representation \cite{MSY}; the remaining problem of converting this
observation into quantum fields associated to Wigner's infinite spin
representation was solved in the cited work. Whereas in that case the string
localization was endemic (every field, including composites has a noncompact
localization), its use for finite helicity representation was only necessary
if one were to use potentials instead of field strengths. The short distance
scale dimension reducing role of string-localization was noted in \cite{MSY}
and the first remarks about how to use this in order to convert
nonrenormalizable pointlike couplings into power-counting renormalizable
interactions of stringlocal fields can be found in \cite{Rio} and \cite{Bros}%
.

Meanwhile there exist strong indications that the noncompact localizable
infinite spin Wigner "stuff" is more than a booster for directing the
conceptual attention to string-localization. Its solely noncompact modular
localization (no pointlike composites) leads to "inert matter"; accepting
the standard picture about counter registration of particles, a particle
event is interpreted as a reduction of a field state into a quasi-local
counter-centered particle wave function\ \cite{Haag}. In case of the
inherently noncompact Wigner stuff the localization of fields and particles
are both noncompact and the situation of a counter registration can not be
realized. However the stability and the coupling to gravitation which are
properties shared by all positive energy representations; for this reason
the Wigner stuff is the ideal candidate of "darkness". In fact interacting
confined matter and free noncompact Wigner stuff stand in a conceptually
interesting contrast \cite{dark}. Although they share the \textit{%
irreducibility} of their zero mass strings (which accounts for the darkness
of free strings and the confinement of interacting strings) whereas the
strings of electrically charged fields of QED remain reducible \cite%
{stringlocal}.

\begin{acknowledgement}
I am indebted to Jens Mund for numerous discussions and I am looking forward
to the publication of his mathematical-conceptual support for the new ideas
as well as projected joint publications. I also thank Karl-Henning Rehren
for a critical reading of the manuscript. Special thanks go to Raymond Stora
who accompanied this project with interest and encouragement and last not
least for sharing his over more than during 5 decades acquired critical
wisdom about gauge theory.
\end{acknowledgement}

\end{document}